\documentclass[prb,aps,twocolumn,10pt,superscriptaddress,notitlepage,longbibliography]{revtex4-1}
\usepackage[margin=1.54cm]{geometry}
\usepackage[caption=false]{subfig}
\usepackage{graphicx}
\usepackage{amsmath,mathtools}
\usepackage{amssymb}
\usepackage{epstopdf}
\usepackage{siunitx}
\usepackage{color}
\usepackage[ngerman,english]{babel}

\usepackage{hyperref}
 \hypersetup{
     colorlinks=true,
     linkcolor=blue,
     filecolor=blue,
     citecolor = magenta,      
     urlcolor=red,
     }

\usepackage{braket}

\usepackage{letltxmacro}
\LetLtxMacro{\ORIGselectlanguage}{\selectlanguage}
\makeatletter
\DeclareRobustCommand{\selectlanguage}[1]{%
  \@ifundefined{alias@\string#1}
    {\ORIGselectlanguage{#1}}
    {\begingroup\edef\x{\endgroup
       \noexpand\ORIGselectlanguage{\@nameuse{alias@#1}}}\x}%
}
\newcommand{\definelanguagealias}[2]{%
  \@namedef{alias@#1}{#2}%
}

\makeatother

\definelanguagealias{en}{english}
\definelanguagealias{EN}{english}
\definelanguagealias{eng}{english}
\definelanguagealias{de}{ngerman}

\begin{document}
	
\title{Dissipative modes, Purcell factors and directional beta factors in gold bowtie nanoantenna structures}

\date{\today}
\author{Chelsea Carlson}
\email{0clac@queensu.ca}
\affiliation{Department of Physics, Engineering Physics and Astronomy,
Queen's University, Kingston, ON K7L 3N6, Canada}
\author{Stephen Hughes}
\email{shughes@queensu.ca}
\affiliation{Department of Physics, Engineering Physics and Astronomy,
Queen's University, Kingston, ON K7L 3N6, Canada}

\begin{abstract}
   We present a detailed quasinormal mode  analysis
   of gold bowtie nanoantennas, and highlight
   the unusual role of the substrate and the onset of multi-mode behaviour. In particular, we show and explain
   why the directional raditiave beta factor is completely dominated by emission into the substrate, and explain how the beta factors and quenching depend on the underlying mode properties. We also quantitatively explain the
  generalized Purcell factors and explore the role of gap size and substrate in detail.
  These rich modal features are essential to understand for future applications
  such as sensing, lasing, and quantum information   processing, for example in the design of efficient single photon emitters.
\end{abstract}

\maketitle

\section{Introduction}
\label{sec:introduction}

The search for efficient quantum light sources
and detectors is a rapidly expanding field in nanophotonics and plasmonics, offering new components in 
chip-based quantum electrodynamics (QED), with applications in quantum sensing and lasing. Part of the goals in achieving such devices, at the level of a single or a few quantum emitters, is to carefully engineer the light-matter interactions at the locations of the quantum emitters (e.g., a quantum dot, defect, or dye-sensitized molecule) to produce large enhancements in the electric field and thus the total spontaneous emission rate via the Purcell effect \cite{purcell_spontaneous_1946}. 

It is well-known that plasmonic structures and antennas offer fundamentally different light-matter coupling  
behavior than dielectric or semiconductor photonics,
e.g.,  they do not suffer from the usual 
spatial diffraction limits on the
effective mode volume and they are intrinsically very fast (offering a broadband response). Consequently, plasmonic structures can enhance light-matter interactions by confining the
optical fields to sub-nanoscale mode volumes, with orders of magnitude smaller mode volumes than the diffraction limit (e.g.,  even ``picocavities'' have been realized \cite{benz_single-molecule_2016})).
Along with ultra-small mode volumes, metallic systems can also produce very high Purcell factors (PFs) -- the enhancement of spontaneous emission relative to the homogeneous background -- reported in the thousands (to hundreds of thousands) both theoretically \cite{akselrod_probing_2014, kamandar_dezfouli_modal_2017, szenes_enhancing_2018} and experimentally \cite{akselrod_probing_2014, wei_ultrahigh_2016}.   However, the PF is not the only figure of merit (FOM) for such systems, since one goal in light emission and sensing is to be able to efficiently read-out the emitted photons
``on demand''. Often this type of FOM is known as either the quantum efficiency or the radiative $\beta$-factor (namely the probability that a single photon emitted will be detected in the far field), the latter of which will be used in this paper. These metrics are also essential to
develop quantum FOMs
for plasmon-based single photon sources~\cite{koenderink_single-photon_2017, Hughes2019}.

% In the context, single photon sources, more than 99\% of on-demand photons must reach the detector in order to achieve high enough efficiency for any quantum information processing. 
Although plasmonic resonators offer high-PF interactions over a large broadband frequency range, which helps with the coupling to 
quantum emitters such as quantum dots and dye molecules \cite{Lyamkina2016,chikkaraddy_single-molecule_2016,luo_plasmon-exciton_2019}, the lossy nature of the metals creates a particularly difficult challenge for the $\beta$-factor, as most of the photons can be lost to Ohmic heating or quenching \cite{khurgin_how_2015, axelrod_hyperbolic_2017}. Yet, high beta factors are required to enable on-demand quantum light sources (and to help detection
and sensitive Raman experiments), and it is important to understand their properties in a detailed an intuitive way.  

Configurations of metallic nano-structures range from flat slab interfaces \cite{zia_geometries_2004} to single or ensembles of nanoparticles \cite{fan_self-assembled_2010} to slot waveguides \cite{guo_optical_2008} to more complex geometries \cite{winkler_direct-write_2017,kamandar_dezfouli_modal_2017,Baranov_novel_2018}, supporting many tunable resonances~\cite{tali_multiresonant_2019} that are highly sensitive to geometry, size, and environment (e.g., background index of refraction or presence of other molecules). These resonator structures support lossy cavity modes  with outgoing boundary conditions (which can be simulated numerically with perfectly matched layers, PMLs) for which the electric field propagates outward to infinity following the Silver-M\"{u}ller radiation condition \cite{kristensen_normalization_2015} (i.e., outgoing/scattered waves must be spherical in the limit of $r\rightarrow\infty$, in the form $\frac{E_0}{r}{\rm e}^{i(kr-\omega t)}$). However,  since the frequency of the resonance is inherently complex, the spatial mode diverges in space  and renders the usual Hermitian normalization conditions invalid since the surface integral of the electric field also diverges at infinity. Thus, the usual normal mode theory cannot be applied to lossy systems. One way to 
overcome this problem, is to use  quasinormal mode (QNM) theory, which
allows a rigorous 
definition of the effective mode volume,
while working directly with the
QNM complex eigenfrequency and thus loss.
The properties of QNMs
 have been shown to accurately describe a plethora light-matter
 interactions of cavity systems \cite{Kristensen2012,bai_efficient_2013,kristensen_modes_2014,kamandar_dezfouli_modal_2017,axelrod_hyperbolic_2017,dezfouli_nonlocal_2017,alpeggiani_quasinormal-mode_2017,lalanne_light_2018}; more importantly, the QNMs allows one to meaningfully talk about cavity FOMs in terms of the underlying mode theory, including coupled modes \cite{vial_coupling_2016,Kristensen2017}.

To achieve small mode volume and high electric field ``hotspots,'' dimer nanoantenna structures (two metal nanoparticles) are often employed with small gap sizes and various geometrical designs, including but not limited to nanorods \cite{Ge2014,shcherbakov_plasmon_2015}, cubes \cite{wu_charge_2016,hu_closely_2019}, disks \cite{alpeggiani_visible_2016}, and nanospheres/shells \cite{liaw_purcell_2009}.
Bowties designs in particular are well exploited~\cite{fromm_gap-dependent_2004, kinkhabwala_large_2009, suh_plasmonic_2012,schraml_optical_2014, chou_chau_tunable_2016, kaniber_surface_2016, Lyamkina2016,yan_rigorous_2018}, though their basic cavity mode structure are not well understood.
Thus there is an important need to better understand
the optical properties of such structures
from a cavity mode perspective, especially their radiative emission properties.

\begin{figure}[thb]
    \centering
    \includegraphics[width=0.75\columnwidth]{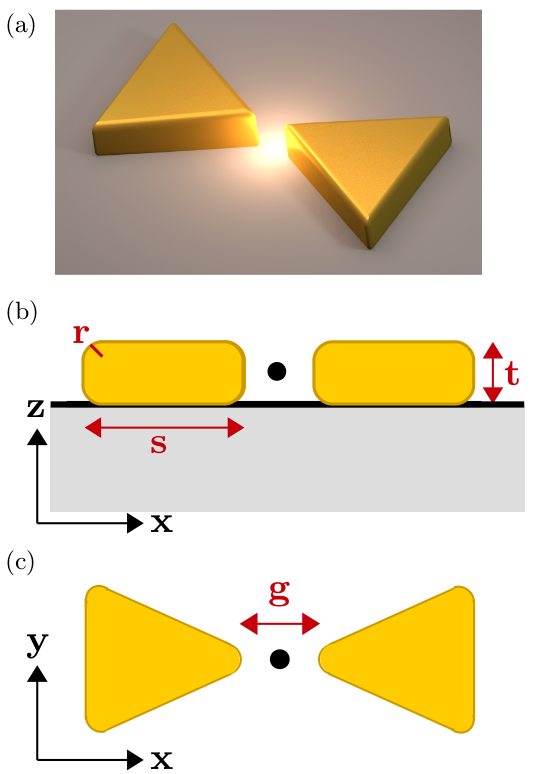}
    \caption{Illustration of a bow-tie dimer on a substrate (a) as well as an $XZ$ and $YZ$ (b and c, respectively) schematic view of the system. The electric dipole is marked by a black marker, and the relevant geometric parameters are labelled in red (see text for values).}
    \label{fig:schematic}
\end{figure}

In this paper, we calculate and study the underlying QNM properties and directional beta factors %COMSOL Multiphysics is used to numerically extract the normalized QNMs 
for a gold bowtie dimer cavities, shown schematically in Fig.~\ref{fig:schematic}. These bowtie geometries are particularly interesting because of their pointed geometry at the gap, which produces a ``lightening-rod effect'' \cite{liao_lightning_1982, gramotnev_gap-plasmon_2012}, higher sensitivity to changes in geometry (i.e., gap, angle of tip, size, etc.)~\cite{fischer_engineering_2008} and polarization, and smaller non-radiative power losses \cite{rogobete_design_2007}. 

%\red{Add a few lines about bowtie antennas being particualtly attractive with a few references, which motivated a QNM analysis of their properties, perhaps also citing teh high beta factor is this is known from other works}

The rest of the paper is organised as follows: in Sec.~\ref{sec:theory}, a brief introduction to QNM theory is provided with all relevant equations used for this study. These modes  are computed using COMSOL Multiphysics and an inverse Green function approach in complex frequency space~\cite{bai_efficient_2013}. 
The spatially dependent PFs
are obtained analytically and also checked
in terms of full dipole simulations, showing excellent agreement.
Section~\ref{sec:results} contains the main results which includes a detailed study of the effect of gap size, dipole position, multi-mode coupling, and substrate index of refraction. The main FOMs we present are the PF and $\beta$-factor, where the mode volume, quality factor, and directionality of the $\beta$-factor are also examined for the first four modes of the dimer systems, over a wide bandwidth. The main results show a splitting of the second-order mode from the inclusion of a substrate, as well as a red-shift of all four modes as the index of refraction of the substrate increases\cite{fischer_engineering_2008,kaniber_surface_2016}. Most notably, a clear directionality to the $\beta$-factor follows an unexpected trend which shows that most of the electromagnetic energy is ``pulled'' into the substrate, rather than reflected upward into the upper hemisphere half-space, as the substrate index increases. These results are then summarized and discussed by comparing the different modes and features in each design. 
% %
% For a dimer gap of 21~nm and no substrate, and a emitter position 5~nm from the bottom of the dimer and in the middle of the gap ($\mathbf{r}$ = [0,~0,~5]~nm), the PF and $\beta$-factor
% for the first (second) mode are approximately 647 (1238) and 83\% (11\%), respectively. 
% %
% With a substrate of refractive index 3.5, the first mode (low energy) of the dimer, with a dipole emitter at an equivalent position, reaches a PF of 629 and a  $\beta$-factor of 74\% -- both larger than the other 3 observed modes. However, more than 95\% of this radiated power is now directed into the substrate (lower hemisphere). 
%
Lastly, in Sec.~\ref{sec:conclusions}, we present our  conclusions. 

%\red{Update with new section - discussion and summary of mode stuff}

\section{Theory}
\label{sec:theory}

Quasinormal modes in optics are mode
solutions of cavity structures with open-boundary cavities and are direct solutions to the vector Helmholtz equation in the complex frequency domain. Assuming  non-magnetic media, 
%then 
the eigenvalue equation for the electric-field
QNM is
\begin{equation}
\boldsymbol{\nabla}\times\boldsymbol{\nabla}\times\tilde{\mathbf{f}}_{{\mu}}\left(\mathbf{r}\right)-\left(\dfrac{\tilde{\omega}_{{\mu}}}{c}\right)^{2}\varepsilon\left(\mathbf{r},\tilde{\omega}_\mu\right)\,\tilde{\mathbf{f}}_{{\mu}}\left(\mathbf{r}\right)=0,
\label{eq:helmholtz}
\end{equation}
where $\tilde{\mathbf{f}}_{{\mu}}$ are the QNMs, $\tilde{\omega}_\mu = \omega_\mu - i\gamma_\mu$  are the complex eigenfrequencies, $\varepsilon$ is the material permittivity (in general, complex), and $\mu$ is the mode number. For the gold dimer, the Drude model of metals is used to describe the permittivity, $\varepsilon_{\rm gold} = 1 - {\omega_p^2}/{\omega(\omega + i\gamma_p)} $, where $\omega_p$ and $\gamma_p$ are the plasma and collision frequencies; for gold, we use 1.26$\times 10^{16}$~rad/s (8.2935~eV) and 0.00141$\times~10^{16}$~rad/s (0.0093~eV), respectively. The photonic Green function, which is the electric field response to a single dipole emitter can be written in terms of a superposition of all QNMs over all pairs of space points, $\mathbf{r}$ and $\mathbf{r}^\prime$\cite{lee_dyadic_1999,Kristensen2017},
\begin{equation}
\mathbf{G}(\mathbf{r}, \mathbf{r}^\prime, \omega) \equiv \sum_\mu A_\mu(\omega)\tilde{\mathbf{f}}_\mu (\mathbf{r}) \tilde{\mathbf{f}}_\mu (\mathbf{r}^\prime),  
\label{eq:2}
\end{equation}
where $A_\mu(\omega)=\frac{\omega}{2(\tilde{\omega}_\mu - \omega)}$ is a complex and frequency dependent coupling constant, while the QNMs 
only depend on
 spatial position.  The
above expansion  is accurate for spatial location within and near the resonator.
In general, QNMs in dispersive media may be
%It is important to note that QNMs ar
inherently non-orthogonal to one another~\cite{perrin_eigen-energy_2016}, but in this work, the coupling of two or more modes are performed under the assumption that the modes can be treated as 
%\textit{approximately} 
orthogonal (i.e., non-diagonal elements of the coupling matrix are negligible)~\cite{bai_efficient_2013,kamandar_dezfouli_modal_2017}. We also check this numerically by comparing with full dipole simulations with no 
%modal 
approximations.

The normalization condition for such modes has been the subject of much interest \cite{Leung1994,Sauvan2013,Ge2014},
since the usual normalization used for {\it normal modes} (formally, with real eigenfrequencies) is no longer valid. 
A simple way to obtain normalized
modes is to exploit the solution
of a scattering problem from
a dipole (at ${\bf r}_0$) in complex frequency space, which yield the mode in normalized
form from the Green function solution \cite{bai_efficient_2013},
%
% For example, for a dielectric system,
% one usually uses $\braket{{\mathbf{f}}_\mu\vert{\mathbf{f}}_\mu} = 1 =  \lim_{V\to\infty}\int_V
% \epsilon(\mathbf{r}) |\mathbf{f}_\mu({\bf r})|^2 d{\bf r}$, which diverges for any open cavity mode with finite dissipation~\cite{Kristensen2012}. The main problem here is that one is using a Hermitian normal-mode theory
% for a non-Hermitian open boundary problem, which in fact
% always yields a complex eigenfrequency solution. Following a single-mode scattering theory approach
%
\begin{equation}
%\begin{aligned}
    \mathbf{\tilde{f}}_\mu(\mathbf{r})  %\frac{\mathbf{E}_\mu(\mathbf{r,\omega})}{\sqrt{A_\mu(\omega)\mathbf{d}\cdot\mathbf{E}_\mu(\%mathbf{r}_0,\omega)}} \\
    = \underbrace{\sqrt{\frac{2i\varepsilon_0(\tilde{\omega}_\mu - \omega)}{\mathbf{J}\cdot\mathbf{E}_\mu(\mathbf{r}_0)}}}_\text{Normalization Constant}{\mathbf{E}_\mu(\mathbf{r,\omega})},
%\end{aligned}    
\label{eq:fr-normalization}
\end{equation}
where $\mathbf{d}$ is the dipole moment (related to the dipole current density, $\mathbf{J}\equiv i\omega\mathbf{d}$), and the electric field of the mode is defined by the Green function and the dipole, $\mathbf{E}(\mathbf{r},\omega) =\frac{1}{\varepsilon_0}\mathbf{G}(\mathbf{r},\mathbf{r}_0,\omega)\cdot\mathbf{d}$, where the position of the dipole is $\mathbf{r}_0$. Note that $\omega$ in this equation is the frequency for which the electric field is obtained which is very close to the pole 
frequency; generally, we use $\omega=(1-10^{-5})\tilde{\omega}_\mu$.
%\blue{you have not introduced $\tilde \omega$ in eq. 3, so this will be confusing to someone who has not doen this before?}

In this formulation, the electric field as all spatial points in the simulation volume is obtained using a dipole current source; we use COMSOL Multiphysics (RF module), a commercially available frequency dependent solver for Maxwell's equations ~\cite{comsol}. There are several ways for which the pole of the modes can be obtained numerically. In this work, the pole is found by iterating through complex frequencies using a Pad\'{e} approximation \cite{bai_efficient_2013}, starting with an initial guess of the pole. For well-conditioned setups %\blue{meaning?} 
in which the mesh and PML settings have been carefully chosen for numerical stability (i.e., so that they do not produce fictitious solutions), and no more than 7-8 iterations are generally needed to converge to the pole frequency of any given mode. This frequency dependent method is not the only way of accurately obtaining the QNM complex poles, and there are also examples that use finite-difference time-domain algorithms (FDTD) \cite{Ge2014, kamandar_dezfouli_regularized_2018},
and other finite-element solvers~\cite{Lalanne2019}.

%\blue{ very very briefly, and say this is one of a few techniques and reference some of teh others, such as our FDTD QNM papers Ge and Hughes Optics Letters, Mohsen and Hughes PRB}

\begin{figure}[thb]
    \centering
    \includegraphics[width=0.75\columnwidth]{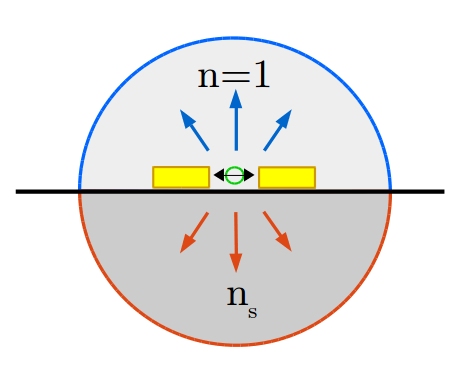}
    \caption{$XZ$ schematic of the entire geometry with the dipole (black arrow), dimer (yellow rectangles), upper hemisphere, lower hemisphere (substrate), and the three surfaces of interest for the Poynting vector -- dipole, far field upper hemisphere and lower hemisphere (green, blue, and red, respectively). }
    \label{fig:setup}
\end{figure}

%\blue{Word of caution - in general for more than one modes you need to account for some nonorthogonaility and solve a matrix~\cite{https://www.osapublishing.org/oe/abstract.cfm?uri=oe-24-24-27137} (also discussed in Sauvan); when Mohsen looked at this before we found such effects negligible, so it migth be wise to say we ignore such effects as we find them to be negligible and refer to this paper}

The complex mode volume, $\tilde V_\mu$ (a ``generalized'' mode volume~\cite{Kristensen2012}), as well as the quality factor, $Q$, are two key FOM for photonic cavities, where $Q_\mu = {\omega_\mu}/{2\gamma_\mu}$. Historally,
in cavity physics, the mode volume is usually discussed in the context of normal modes. However, due to the inherent losses built into the QNM, the mode volume is expressed as a complex quantity, defined by:
\begin{equation}
    \tilde V_\mu(\mathbf{r}) = \frac{1}{\varepsilon_{\rm B} \tilde{\mathbf{f}}^2_\mu(\mathbf{r})},  
    \label{eq:veff}
\end{equation}
where the \textit{effective} mode volume
(e.g., for use in Purcell's formula) is defined $V_{\rm eff}(\mathbf{r})\equiv {\rm Re}[V(\mathbf{r})]$, and it is also a spatially dependent quantity;
and $\varepsilon_{\rm B}$ is the background
dielectric constant where the dipole is embedded
(i.e., $n^2$ or $n_{\rm s}^2$ as shown in 
Fig.~\ref{fig:setup}).

The generalized Purcell factor is given by the ratio of the spontaneous emission rate
with the resonator structure
relative to the homogeneous medium~\cite{kristensen_modes_2014}:
\begin{align}
\label{QNMpurcell}
\begin{split}
    F_{{\rm P}}^{\rm QNM}({\bf r}_0,\omega) &=\frac{\Gamma(\mathbf{r}_{0},\omega)}{\Gamma_{0}(\mathbf{r}_{0},\omega)}
    =\frac{\mathbf{n}_{\rm d}\cdot{\rm Im}\{\mathbf{G}\left(\mathbf{r}_0,\mathbf{r}_0,\omega\right)\}\cdot\mathbf{n}_{\rm d}}{\mathbf{n}_{\rm d}\cdot{\rm Im}\{\mathbf{G}_{\rm hom}\left(\mathbf{r}_0,\mathbf{r}_0,\omega\right)\}\cdot\mathbf{n}_{\rm d}}\\
    &=1+\frac{6\pi c^{3}}{\omega^{3}n_{\rm B}}\,\mathbf{n}_{\rm d}\cdot{\rm Im}\{\mathbf{G}_{\rm c}\left(\mathbf{r}_0,\mathbf{r}_0,\omega\right)\}\cdot\mathbf{n}_{\rm d},
\end{split}
\end{align}
where ${\bf n}_{\rm d}$ is the dipole direction.
% and $\epsilon_{\rm B}$ is the background
% dielectric constant where the dipole is embedded
% (i.e., $n^2$ or $n_{\rm s}^2$ as shown in 
% Fig.~\ref{fig:setup}).
% The factor of 1 is for
% dipole
% locations outside the scattering geometry
% Note that  we have added the extra factor of $1$, which can be derived from from a Dyson equation scattering problem for dipole located outside the resonator (essentially the contribution from the homogeneous radiation modes) \cite{ge_quasinormal_2014}. The actual QNM contribution here is thus the modification to unity.

It is important to also check the validity of a single or few mode 
QNM model. To do this, we can also obtain the numerical full-dipofrom a dipole at position
${\bf r}_d$,
can be numerically obtained 
%in COMSOL 
by obtaining the surface-integrated Poynting vector, from
\begin{equation}
    F_p^{\rm num}(\mathbf{r}_0) = \frac{\int_s \hat{\mathbf{n}}\cdot\mathbf{S}_{\rm dipole, total}(\mathbf{r};\omega) \rm{d}A}{\int_{s} \hat{\mathbf{n}}\cdot\mathbf{S}_{\rm dipole, background}(\mathbf{r};\omega) \rm{d}A}, 
    \label{eq:comsolFP}
\end{equation}
where $\mathbf{S}(\mathbf{r},\omega)$ is the Poynting vector along the surface $s$ of a small sphere centred around the  {\it finite-size} dipole ($\sim$\,1-nm radius) at $\mathbf{r}_0$ (shown in green in Fig.~\ref{fig:setup}).

Using a similar approach, we can also obtain
the numerical beta factor,
\begin{equation}
    \beta(\mathbf{r}_0) = \frac{\int_{s^\prime} \hat{\mathbf{n}}\cdot\mathbf{S}_{\rm PML, total}(\mathbf{r}^\prime;\omega) \rm{d}A^\prime}{\int_{s} \hat{\mathbf{n}}\cdot\mathbf{S}_{\rm dipole, total}(\mathbf{r};\omega) \rm{d}A},
    \label{eq:comsolbeta}
\end{equation}
where $\mathbf{S}(\mathbf{r}^\prime,\omega)$ is the Poynting vector along the surface $s^\prime$ over the spherical boundary between the system and the first PML, and $\hat{\mathbf{n}}$ is the unit vector normal to the integration surface; this allows us to capture the power {\it in} and {\it out} of the system to obtain meaningful $\beta$ and $F_p$ quantities. 
For a lossless dielectric system, then
the beta factor is unity. Thus it had the interpretation
of the probability that a single emitted photon will be emitted radiatively from the entire
resonator system.
%  The reason why the Poynting vector is calculated over a small sphere around the dipole and not simply \textit{at} the location of the dipole is purely for numeric stability; if there are small issues with the meshing of the system, then a single point measurement is not as reliable as that of the average or integral around that point. 

For the directional
dependence of the emission,
 we adopt a convention for the $\beta$-factor which splits  into two quantities: the normalized radiated power in the +$z$ direction and -$z$ direction, which are shown in blue and red, respectively, in Fig.~\ref{fig:setup}. These quantities are defined as: 
\begin{equation}
    \beta^{+}(\mathbf{r}_0) =  \frac{\int_{s\prime+} \hat{\mathbf{n}}\cdot\mathbf{S}_{\rm PML, total}(\mathbf{r}^\prime;\omega) \rm{d}A^\prime}{\int_s \hat{\mathbf{n}}\cdot\mathbf{S}_{\rm dipole, total}(\mathbf{r};\omega) \rm{d}A},
    \label{eq:comsolbeta+}
\end{equation}
\noindent and,
\begin{equation}
    \beta^{-}(\mathbf{r}_0) = \frac{\int_{s^\prime-} \hat{\mathbf{n}}\cdot\mathbf{S}_{\rm PML, total}(\mathbf{r};\omega) \rm{d}A}{\int_s \hat{\mathbf{n}}\cdot\mathbf{S}_{\rm dipole, total}(\mathbf{r};\omega) \rm{d}A},
    \label{eq:comsolbeta-}
\end{equation}
 respectively, where $s^\prime+$($s^\prime-$) is the surface of the top(bottom) hemisphere of the PML boundary.

\section{Results}
\label{sec:results}

For our numerical calculations, the gold bowtie dimers are chosen to follow typical experimental dimensions reported in the literature\cite{Lyamkina2016,Schraml2016,kaniber_surface_2016}, including edges and vertices that are rounded. It is important to note that sharp features in metals are also notoriously difficult to model in any finite-mesh solving software, with the potential to create spurious hot-spots in the electric field (numerical artifacts). Thus, 
all edges and vertices are rounded with a curvature of $r$=3~nm. The other dimensions, labelled in Fig.~\ref{fig:schematic}, are set to $s$=90~nm and $t$=35~nm, and the triangles are equilateral. The gap, $g$, is defined as the gap between the un-rounded vertices, so for the geometry of equilateral triangles, the \textit{true} gap size is larger by 2$r$, such that $g\rightarrow g+2r$ (i.e. for a gap of 15~nm, the true gap is 21~nm). For clarity, only the true gap will be referred to here.
%for numerical stability, as well as consistency with experimental designs, 

\begin{figure}[htb]
    \centering
    \includegraphics[ width=1\columnwidth]{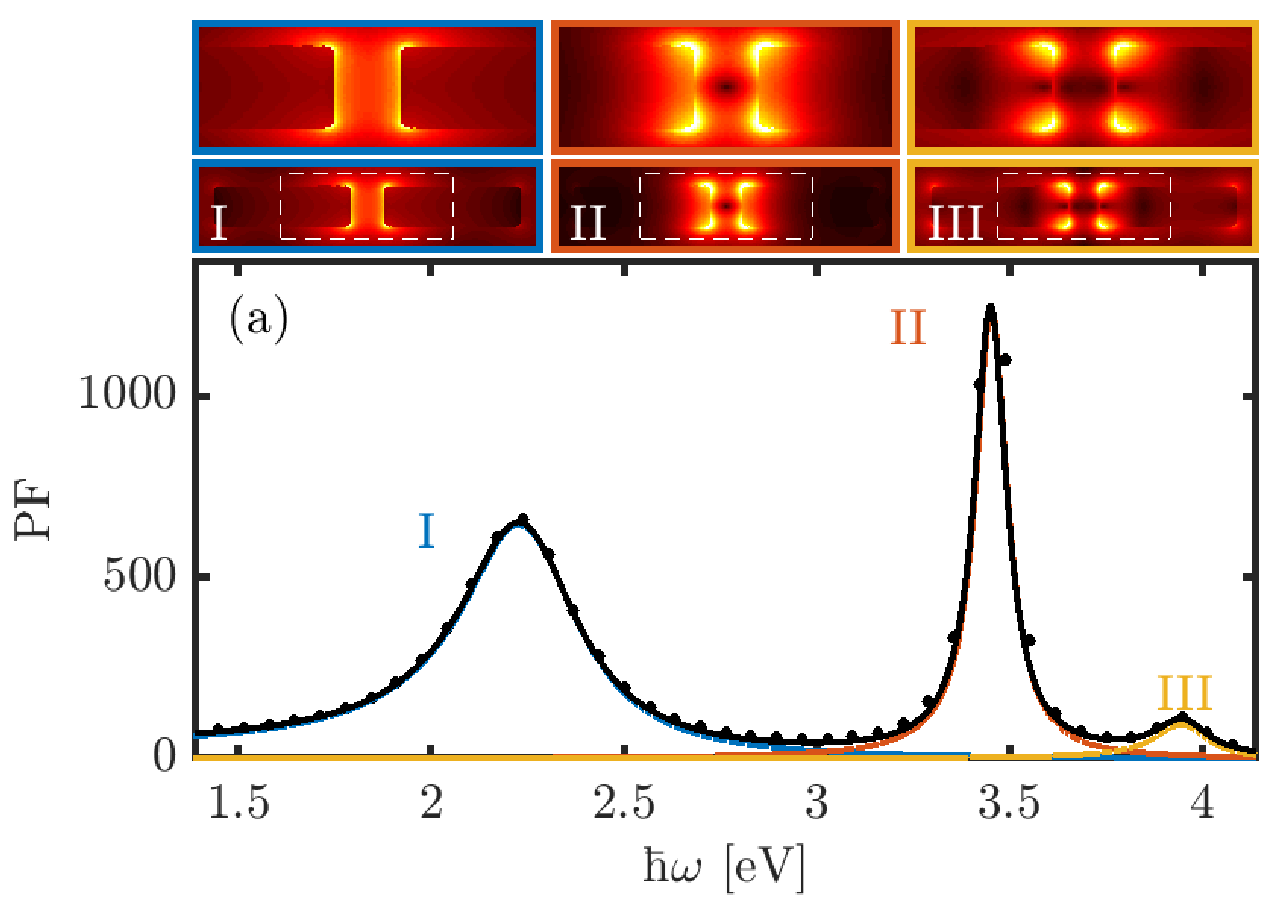}
    \includegraphics[ width=1\columnwidth]{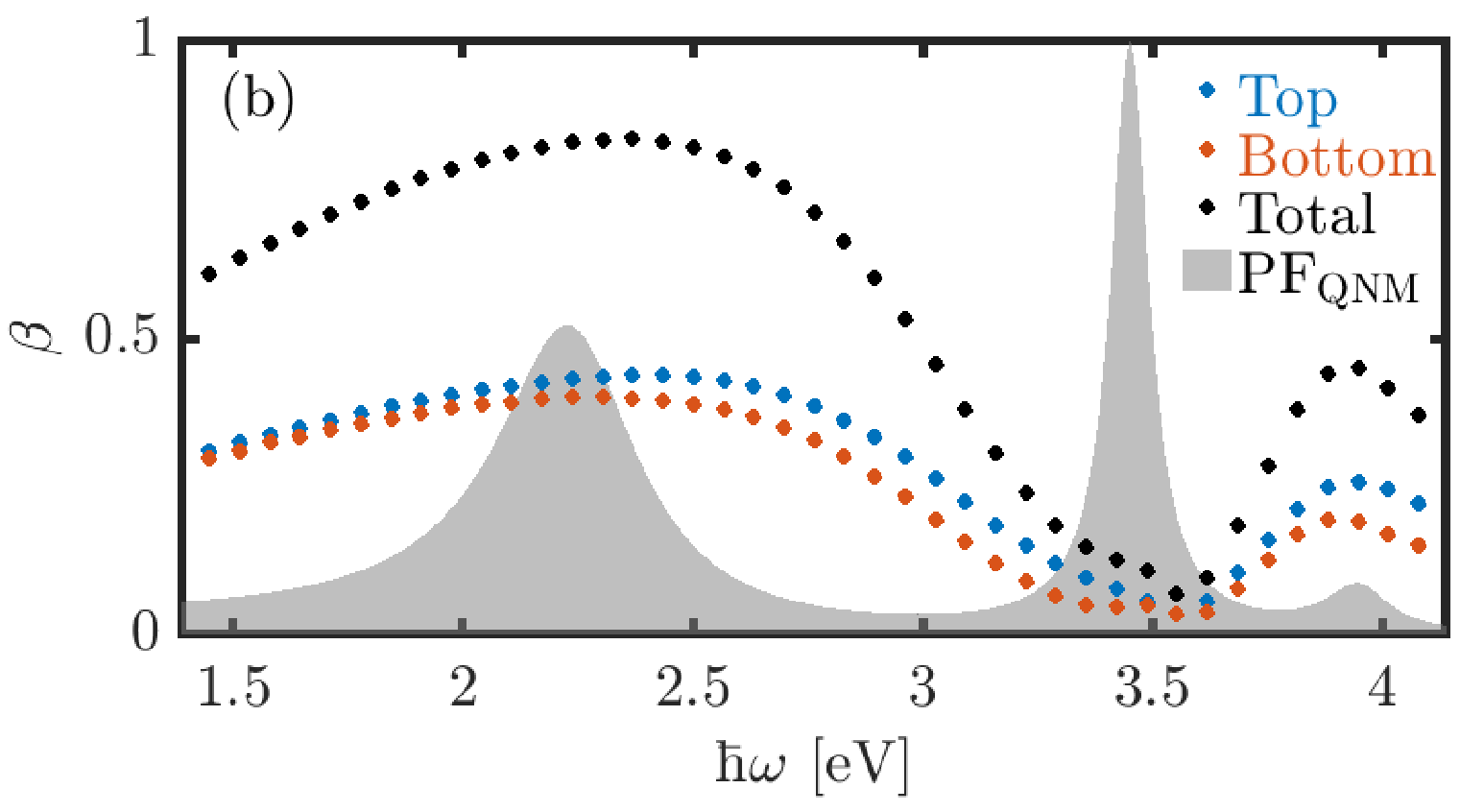}
    \caption{No substrate case ($n_{\rm s}$=1). The PF in the $x$ direction and $\beta$-factor for a dimer with a gap of 21~nm and no substrate ($n_s$=1). The position of the dipole is set to the center of the gap ($x$=0, $y$=0) 5~nm above the substrate boundary ($z$=5~nm). Full-dipole calculations are given by markers, single QNM PF calculations are given by coloured lines, and the multi-mode (3 modes) QNM PF is given by a black line. The QNM field $XZ$ profiles ($\vert\tilde{\mathbf{f}}_\mu(\mathbf{r})\vert^{0.5}$ in arbitrary units) at $y$=0 for each mode are given at their respective pole frequencies, as well as a zoomed-in profile in the gap of the dimer. %\blue{I don't think I would show ana dat apoint above 4.1 eV as there is no P factor - so truncate the max to near there; fonts also a tad small here - axis labels }
    }
    \label{fig:ns_1}
\end{figure}
\begin{figure}[htb]
    \centering
    \includegraphics[trim = 0cm 0cm 0cm 0cm, clip=true, width=\columnwidth]{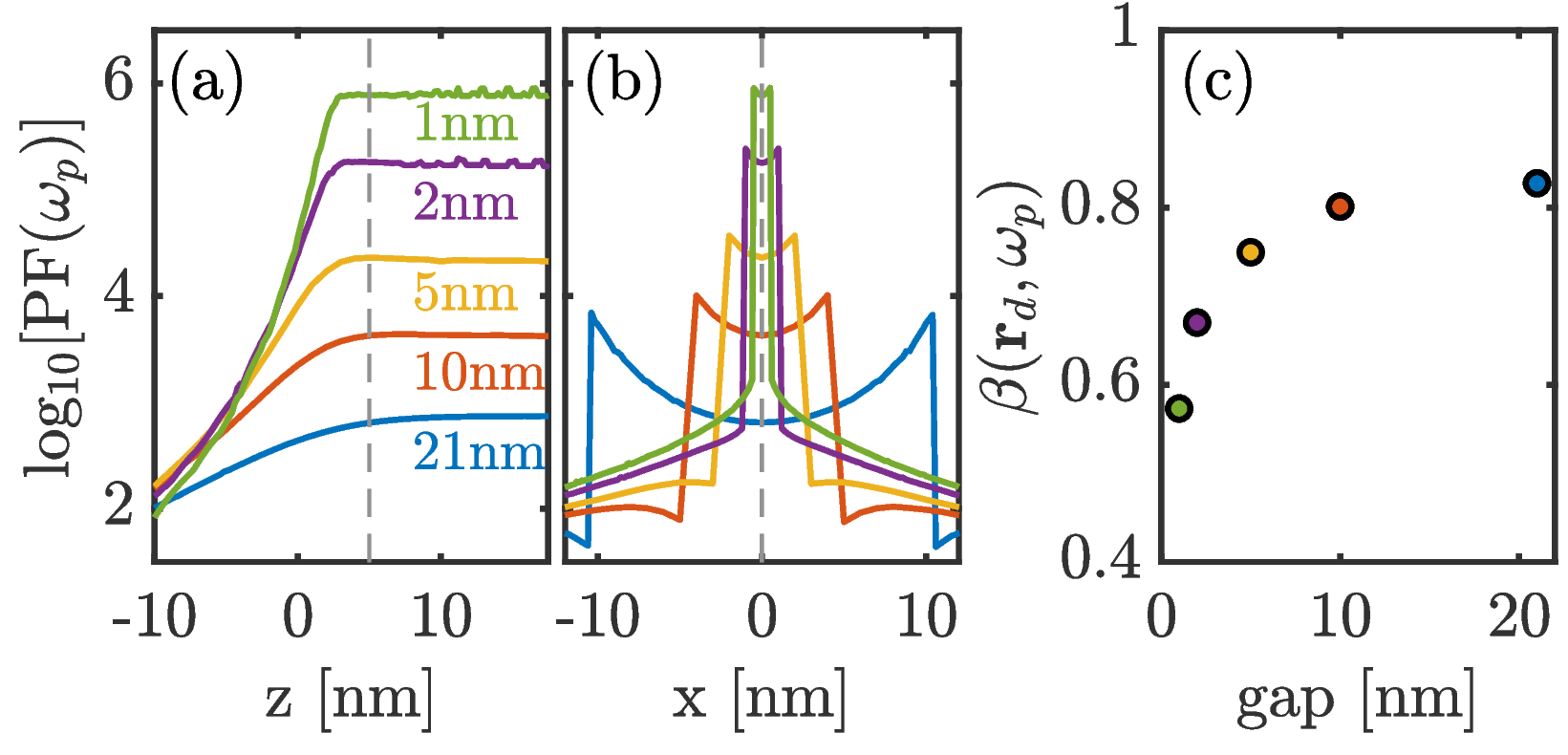}
    \caption{No substrate case 
    ($n_{\rm s}$=1. A summary of the (a-b) PF (as log$_{10}$(PF)) %\blue{1-2 nm cases look a bit choppy - still needs commented on in text - is this real or numerics? are teh x and y labels large enough fonts - no smaller than figure caption text?} addressed in text now -- it is numerics due to the conversion from comsol to matlab meshing
    and (c) $\beta$-factor of mode I for $n_s$=1 at the pole frequency (real part of $\tilde{\omega}_p$) as a function of gap size and dipole position in (a) $z$ (where $z$=0 is the substrate boundary) and (b) $x$. Grey dashed lines indicate the dipole location of the simulation as well as the location of the $\beta$-factor ($\mathbf{r}_d$) (evaluated at the mode resonance peak).  }
    \label{fig:gapanalysis}
\end{figure}

\begin{figure*}[htb]
    \centering
    \includegraphics[ width=1\columnwidth]{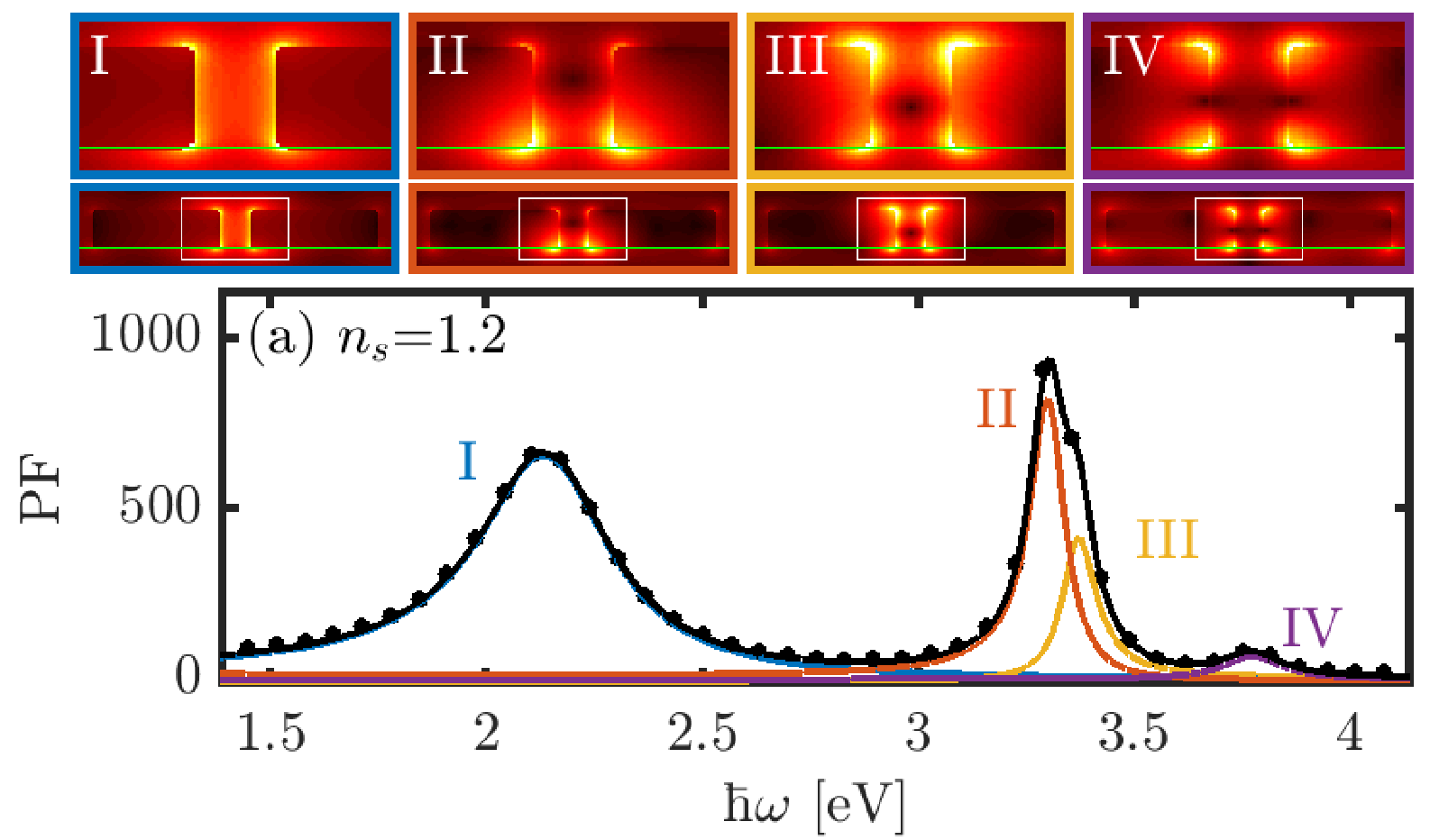}
    \includegraphics[ width=1\columnwidth]{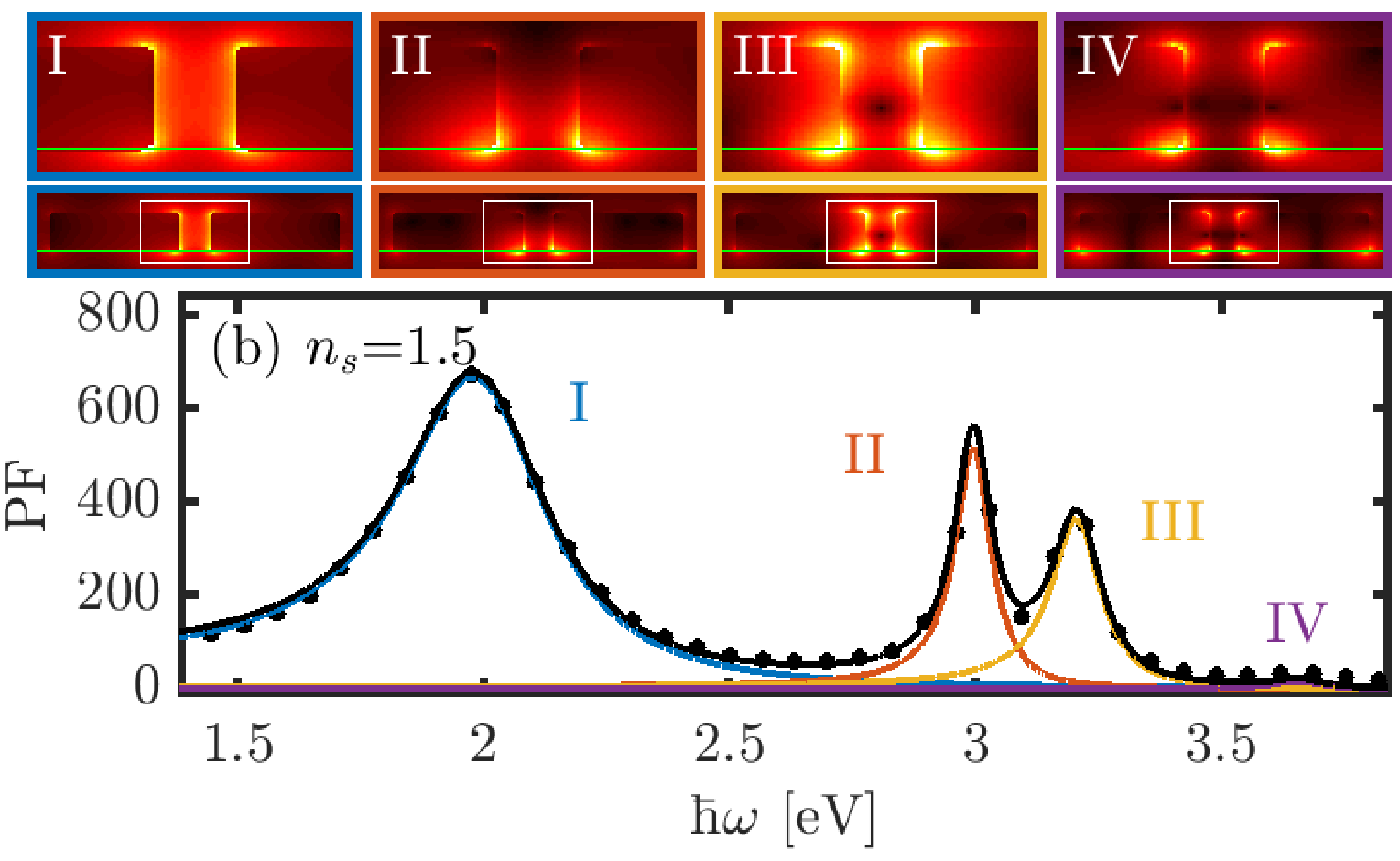}
    \includegraphics[ width=1\columnwidth]{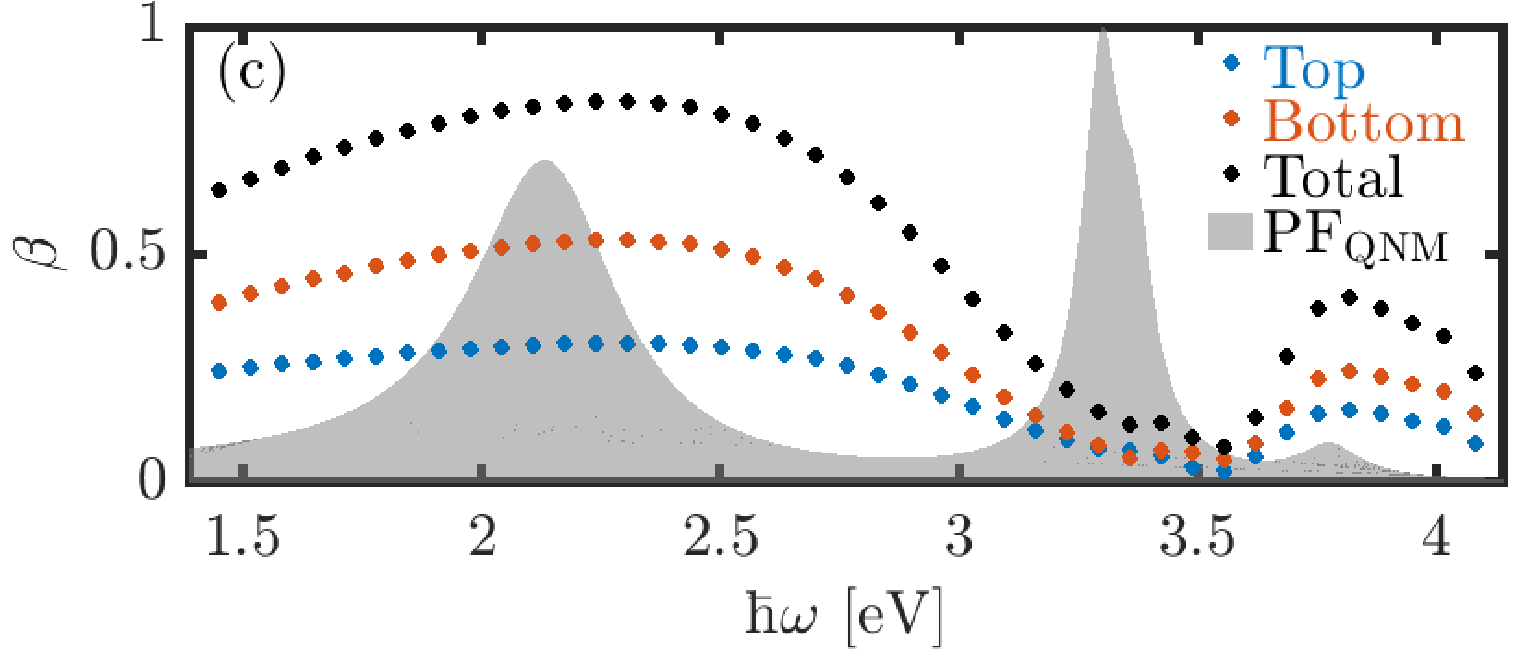}
    \includegraphics[ width=1\columnwidth]{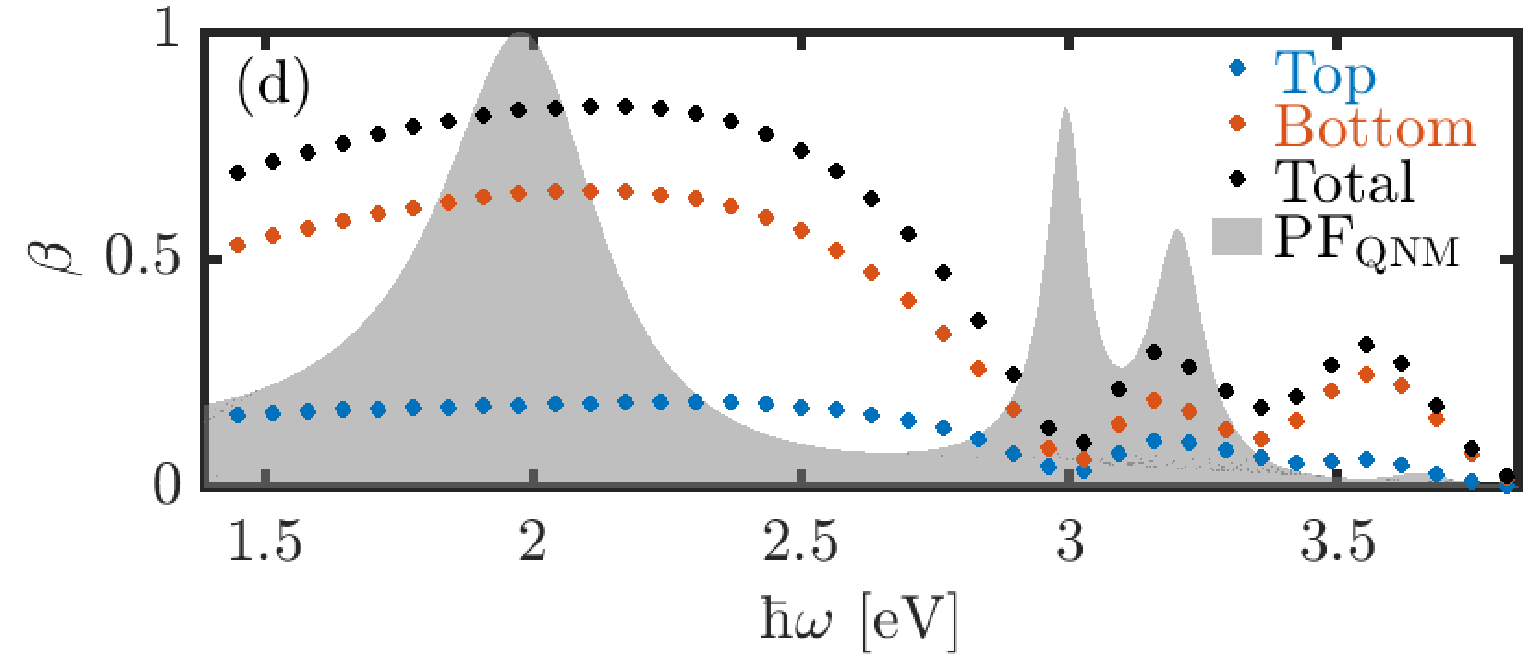}
    \vspace{-0.3cm}
    \caption{$n_{\rm s}$=1.2 (left) and $n_{\rm s}=$1.5 (right) substrate. The PF (a-b) and $\beta$-factor (c-d) for a dimer with a gap of 21~nm and a low-index substrate of $n_s=$1.2 (a,c) and $n_s=$1.5 (b,d). The position of the dipole is set to the center of the gap ($x$=0, $y$=0) 5~nm above the substrate boundary ($z$=5~nm). Full-dipole calculations are given by markers, single QNM PF calculations are given by coloured lines, and the multi-mode (4 modes) QNM PF is given by a black line. The QNM field $XZ$ profiles ($\vert\tilde{\mathbf{f}}_\mu(\mathbf{r})\vert^{0.5}$ in arbitrary units) at $y$=0 for each mode are given at their respective pole frequencies, as well as a zoomed-in profile in the gap of the dimer. %\blue{Again, I don't think I would show any data points above 4. eV in teh left one and around 3.4 in teh rigth one as there is no P factor - so truncate the max to near there; in (b) I find it hard to see the data points (could be a little bigger), and especially in 
    %\blue{(d) why to they appear to be clumped in 2s, unlike teh ones in (a) and (c) - why are these not evenly spaced here? } \red{When I calculated them, I messed up the original spacing of omega, resulting in this. Is it too distracting?}
    }
    \label{fig:ns_low}
\end{figure*}

\subsection{No substrate dimers ($n_{\rm s}$=1)}
\label{sec:nosub}

First, the PF and $\beta$-factor are examined for the simplest case of no substrate ($n_{\rm s}$=1) with a dimer gap of 21~nm, shown in Fig.~\ref{fig:ns_1}. Here, markers indicate the full-dipole values obtained using Eqs.~\eqref{eq:comsolFP}-\eqref{eq:comsolbeta-}, and the QNM calculations for the first three modes (I-III) are given by solid colored lines [\eqref{QNMpurcell}]. 
 To better see the correlation with PF and beta factor, the QNM-calculated PF is shown in arbitrary units for the beta simulation. There are a few notable features on this graph, starting with the excellent agreement between the full-dipole PF and the 3-mode QNM calculation over the entire frequency range shown (spanning around 3 eV). The PF of mode I and II exceed 600 and 1200, respectively. Interestingly, the $\beta$-factor exceeds 80\% and 40\% for mode I and III but is greatly suppressed below 15\% for mode II, which means that although mode II has the largest PF of the three modes, it is mainly non-propagating to the far-field. Lastly, there is some asymmetry between the upper and lower hemispheres, even though there is no substrate to break the symmetry, but it is expected that the upper hemisphere will have a slightly larger $\beta$-factor due to the off-center geometry of the COMSOL setup (see Fig.~\ref{fig:setup}) and location of the dipole. The largest absolute difference is approximately 5\%.

We also note that the nonraditive modal contribution to the decay can also be obtained directly from the
QNM technique, through:~\cite{anger_enhancement_2006}% \blue{check this is the correct ref and also ref the Novotny paper that we usually reference for this}
\begin{equation}
    \Gamma_{\rm NR}(\mathbf{r}_d,\omega) = \frac{2}{\hbar\omega\varepsilon_0}\int_V {\rm Re}[\mathbf{j}(\mathbf{r},\omega)\cdot \mathbf{G}^*(\mathbf{r},\mathbf{r}_d, \omega)\cdot \mathbf{d}]{\rm d}\mathbf{r},
\end{equation}
where $\mathbf{j}(\mathbf{r},\omega) = \omega {\rm Im}[\varepsilon(\mathbf{r},\omega)] \mathbf{G}(\mathbf{r},\mathbf{r}_d,\omega)\cdot \mathbf{d}$ is the induced current
density within the metal dimer.
In this way,
$\beta = 1 - \frac{\Gamma_{\rm NR}}{\Gamma}$ where $\Gamma$ is the total decay rate (radiative + non-radiative, or PF$\times\Gamma_0$).
Note that the Green functions can be obtained directly from Eq.~\eqref{eq:2}.
However, this quantity can be difficult to
obtain numerically (requiring a complex 3D integration per frequency), and it is easier
to just obtain this directly from the dipole simulation. Nevertheless, this expression is useful to explain the general quenching of mode III, since the non-radiative decay is directly proportional to $\vert \mathbf{f}_\mu\vert^2$ within the metal.

%\blue{You need to discuss the results in fig 4 before comign to 5, unless I missed it?}
Figure.~\ref{fig:gapanalysis} shows the peak PF of mode I (see Fig.~\ref{fig:ns_1}) as a function of gap size and $z$ position and $x$ position, as well as the $\beta$-factor evaluated at the dipole location ($z$=5~nm and $x=y=$0~nm). As expected, the PF increases as the dipole location gets closer to the metal boundaries at the cost of the $\beta$-factor decreasing. The small oscilaltions in the PF, particularly seen for the gap of 1 and 2~nm, are numerical artifacts/fluctuations arising from difficulty interpolating the spatial mesh at such a small scales.
%with finite computational resources.   

\subsection{Dimers on a low-index substrate}
\label{sec:lowsub}

\begin{figure}[thb]
    \centering
    \includegraphics[ width=1\columnwidth]{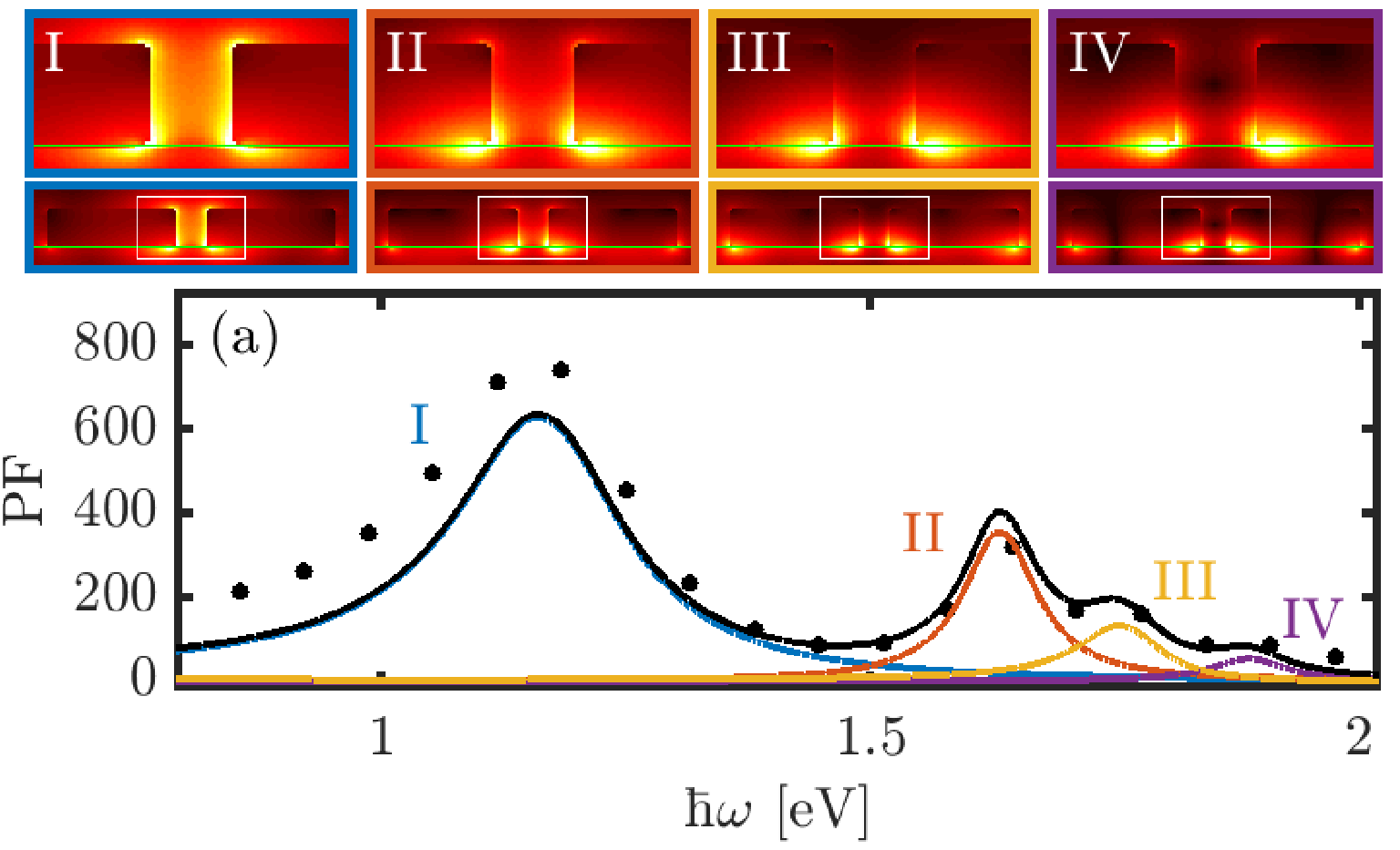}
    \includegraphics[ width=1\columnwidth]{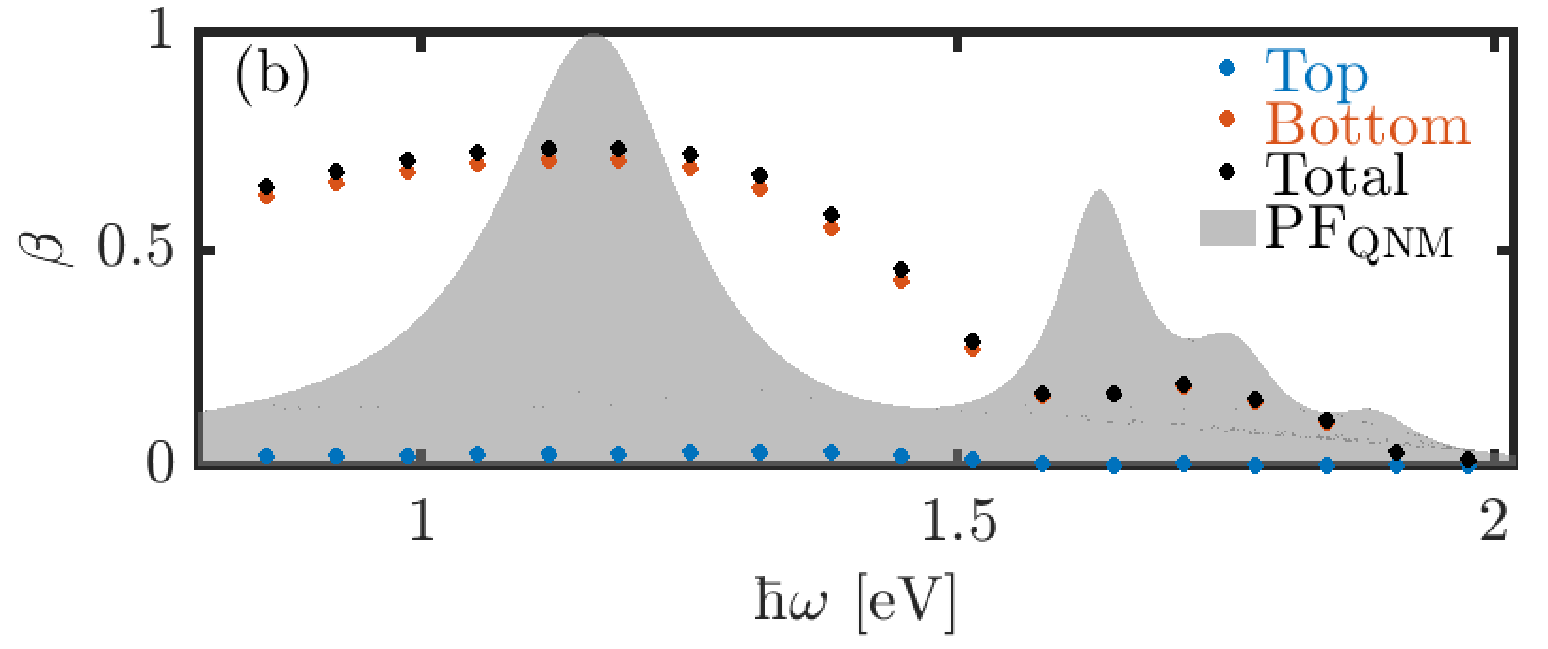}
    \caption{$n_{\rm s}$=3.5 substrate. The PF  and $\beta$-factor for a dimer with a gap of 21~nm and a high-index substrate. The position of the dipole is set to the center of the gap ($x$=0, $y$=0) 5~nm above the substrate boundary ($z$=5~nm). Full-dipole calculations are given by markers, single QNM PF calculations are given by coloured lines, and the multi-mode (3 modes) QNM PF is given by a black line. The QNM field $XZ$ profiles ($\vert\tilde{\mathbf{f}}_\mu(\mathbf{r})\vert^{0.5}$ in arbitrary units) at $y$=0 for each mode are given at their respective pole frequencies, as well as a zoomed-in profile in the gap of the dimer. %\blue{I don't think I would show ana dat apoint above 2. eV as there is no P factor - so truncate the max to near there} 
    }
    \label{fig:ns_3.5}
\end{figure}

\begin{figure*}[thb]
    \centering
    \includegraphics[trim = 0cm 0cm 0cm 0cm, clip=true, width=1\columnwidth]{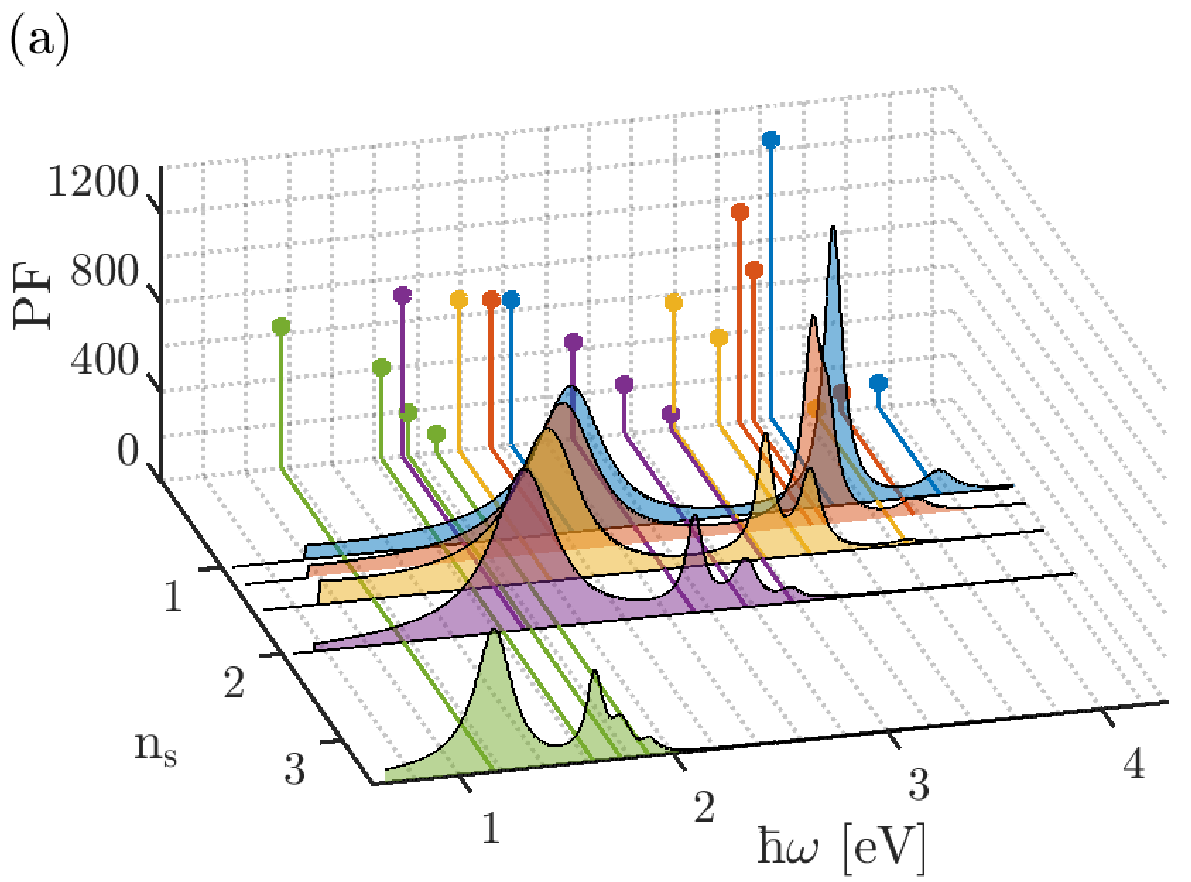}
    \includegraphics[trim = 0cm 0cm 0cm 0cm, clip=true, width=1\columnwidth]{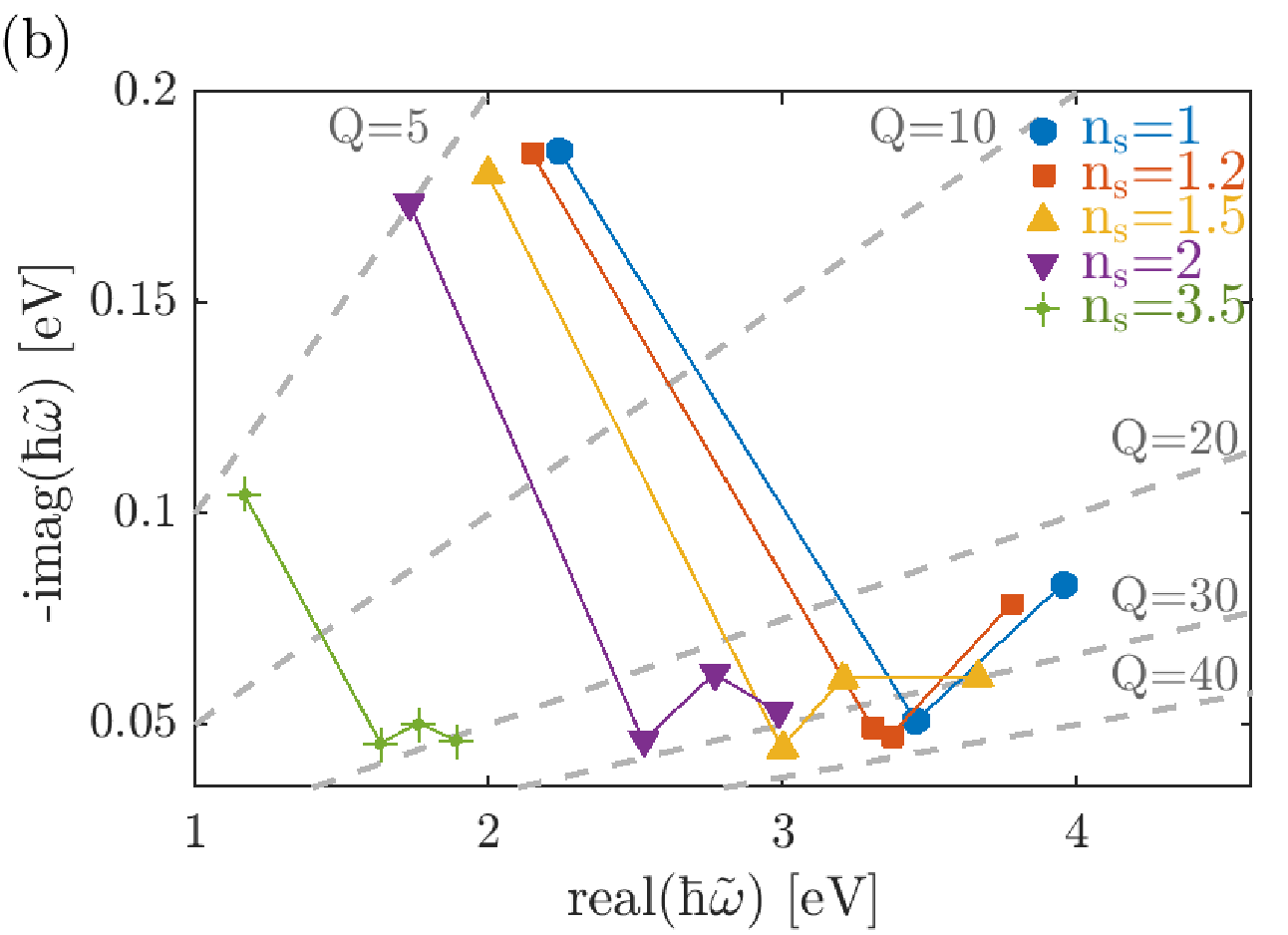}
    \caption{(a) Summary of the PFs obtained from the QNMs, showing the peak PF for each mode as a function of substrate index of refraction. (b) Summary of of the complex poles for each mode as a function of substrate index of refraction. Constant dashed lines show quality factor, and although the modes are discrete, connecting solid lines between data points are shown to guide the eye. 
    %\blue{check especially axis labels and other fonts are large enough ($n_s$ for example is hard to see for me) - namely no smaller than fig caption fonts}
    }
    \label{fig:ns_summary}
\end{figure*}

In the majority of experimental dimer setups, the dimers are placed on top of a substrate with an index of refraction greater than 1, usually glass or a high-index semiconductor such as GaAs \cite{chen_metallodielectric_2012,schraml_optical_2014,Lyamkina2016,kaniber_surface_2016}. To study the effects of adding a high-index substrate, the QNMs are examined starting at $n_s$=1.2, increasing to 1.5 (glass), 2.0, and then finally to 3.5 (e.g., GaAs).

Figure~\ref{fig:ns_low} shows the same simulations as in Fig.~\ref{fig:ns_1} except we now use  $n_{\rm s}$ = 1.2 and 1.5. Immediately upon breaking the spatial symmetry (in the vertical direction), as seen by the $n_{\rm s}$=1.2 results, mode II splits into two distinct modes (now labelled by II and III) but mode I and III (now
labelled IV) remain unchanged apart from minor perturbations from symmetry breaking. This splitting is spectrally more dramatic as the index of the substrate increases, and one can see that this also prompts mode I to have the largest PF of the four modes. Also, between mode II and III, only one of them (II) is suppressed in $\beta$ while the other (III) spikes back up, confirming further that these are two distinctly different modes. 

It is also important to note the effect of the substrate on the symmetry of the 
direction $\beta$-factor; against intuition with a dipole above a slab of dielectric, which is that the power flow from a dipole should be directed mainly into its own hemisphere in the presence of a substrate (even a semiconductor which is a very poor mirror), it is seen that the opposite trend is present. Namely, as the index of the substrate increases, the power flow into the lower (substrate) hemisphere becomes much larger relative to the upper hemisphere emission. This can be  explained by examining the spatial mode profiles of the QNMs. There, the electric field is seemingly ``pulled'' downward into the substrate, so if the highest points of the field lives below the surface of the substrate, then it is in-fact directed downward. This result is important,
as typical experiments would measurement the $\beta$-factor
from the power flow in the reflected (upper) direction, and thus there is a lot of power flow that is potentially missed by not considering the power flow in the downward direction.   

\subsection{Dimers on a high-index substrate}
\label{sec:highsub}

\begin{table}[]
    \centering
    \begin{tabular}{|c|c|c|c|c|}
         \multicolumn{5}{c}{$\rm n_{\rm s} = 1.0$} \\
         \hline
         \multicolumn{1}{|c}{Mode} & \multicolumn{1}{c}{I} & \multicolumn{2}{c}{II} & \multicolumn{1}{c|}{III}\\
         \hline
         $\rm \hbar\tilde{\omega}_p~[eV]$  &  2.23908 &  \multicolumn{2}{|c|}{3.45248} &  3.95269 \\
          & - $i$0.18611 &  \multicolumn{2}{|c|}{- $i$0.05084} &  - $i$0.08333 \\
         $\rm Q$ & 6.0  &\multicolumn{2}{|c|}{34.0}  & 23.7  \\
         $\rm  V_{eff}(\mathbf{r}_0)/(\frac{\lambda_0}{n})^3$&  0.00071   &\multicolumn{2}{|c|}{ 0.00209}  &    0.01902 \\
         $\rm PF(\mathbf{r}_0)$ & 646.6 &\multicolumn{2}{|c|}{1238.3}  & 94.8  \\
         $\rm \beta_{up}(\mathbf{r}_0)$& 0.43 &\multicolumn{2}{|c|}{0.06}  & 0.26 \\
         $\rm \beta_{down}(\mathbf{r}_0)$& 0.40 &\multicolumn{2}{|c|}{0.05}  & 0.19 \\
         \hline
         \multicolumn{5}{c}{ ~ }\\
         \multicolumn{5}{c}{${\rm n_s = 3.5}$} \\
         \hline
         \multicolumn{1}{|c}{Mode} & \multicolumn{1}{c}{I} & \multicolumn{1}{c}{II} & \multicolumn{1}{c}{III} & \multicolumn{1}{c|}{IV} \\
         \hline
         $\rm \hbar\tilde{\omega}_p~[eV]$  & ~1.16635~ & ~1.63159~ & ~1.75944~ & ~1.88932~\\
          & - $i$0.10463~ & ~- $i$0.04555~ & ~- $i$0.05040~ & ~- $i$0.04623~\\
         $\rm Q$ & 5.6 & 17.9 & 17.5 & 20.4  \\
         $\rm  V_{eff}(\mathbf{r}_0)/(\frac{\lambda_0}{n})^3$& 0.00067 & 0.00381 & 0.00988 & 0.02859 \\
         $\rm PF(\mathbf{r}_0)$ & 628.7 & 356.4 & 133.8 & 54.5  \\
         $\rm \beta_{up}(\mathbf{r}_0)$& 0.03 & 0.002 & 0.004  & 0.002 \\
         $\rm \beta_{down}(\mathbf{r}_0)$& 0.71 & 0.17 & 0.17 & 0.04 \\
         \hline
    \end{tabular}
    \vspace{-0.2cm}
    \caption{Summary table for $n_{\rm s}$ = 1 and 3.5, where the pole frequency, quality factor, mode volume, Purcell factor, and up/down $\beta$-factors are given. The location, $\mathbf{r}_0$, is set at the dipole location of [0, 0, 5]~nm ($n(\mathbf{r}_0)$ = 1) and $\lambda_0 =2\pi c/$Re$[\tilde{\omega}_p]$.}% \blue{I would make the grey and light green text labels a little darker - the $Q=5$ etc - but no need to change the lines, but the text should be clear for the reader}}
    \label{tab:summary}
\end{table}

Next, the same analysis is applied to the larger index of $n_{\rm s}$=3.5, which is similar to GaAs in the optical regime of the spectrum, as seen in Fig.~\ref{fig:ns_3.5}. As expected from the low-index data, there is a significant red-shift of the four modes of interest, and the $\beta$-factor is now almost completely dominated by emission into the lower hemisphere. Also shown in Fig.~\ref{fig:ns_3.5}, are the mode profiles of the $x$ and $y$ components of the electric field for each of the four modes at their pole frequencies, showing the different symmetries of the different modes. 
However, we note that the QNM fit to the full-dipole calculations is not as good as in the other data sets, with a larger
discrepancy for lower frequencies. This suggests that the quasi-static contributions of the Green function \cite{Ge2014njp} could be playing a a more important role here, which we discuss in more detail in the next section.
Alternatively, there may be some
background small $Q$ modes. Nevetheless, we still
obtain a very good QNM fit to all but the lowest
mode.

Figure~\ref{fig:ns_summary} shows a summary of the QNM PFs as a function of the substrate index of refraction, highlighting the peak PFs for each mode at its pole frequency. This illustrates the red-shifting of all modes as well as the splitting of the original mode II into two distinct modes. In addition, Fig.~\ref{fig:ns_summary} provides a summary of the QNM pole frequencies in complex frequency space. The dashed lines represent constant quality factors, and the lines connecting markers are simply used to guide the eye and group data points for each index of refraction (the modes are, indeed, discrete from each other).    

Table~\ref{tab:summary} summarizes the PF, $\beta$-factor (upper/lower), pole frequency, quality factor, and effective mode volume for all of the calculated modes for $n_s=1$ and $n=3.5$. The upper $\beta$-factor for the high-index substrate is approximately 2-5\% of that in the lower hemisphere with the lower $\beta$-factor reaching as high as 0.71 for mode I. Compared to no substrate, which has a peak $\beta$-factor \textit{total} of 0.83 split evenly between upper and lower hemispheres, the directional behaviour of $\beta$ is important to consider when measuring the quantity. The mode volumes, quality factors, and PFs for each case are similar in magnitude, however the higher index appears to result in over-all larger mode volumes as well as lower quality factors and lower PFs -- especially for the special case of mode II from no substrate, which `splits' into two lower Q modes (II and III) with substrate where the Q of these modes is approximately half that of the original single mode.

\subsection{Discussion and Graphical Summary of Mode Contributions for Different Substrates}
\label{sec:discussion}

To compare briefly to other related works in the literature, a 2012 study by Chen \textit{et al.}\cite{chen_metallodielectric_2012} showed that for a single metal nanoparticle on a substrate, the role of the index of the substrate was counter to our results; namely, they calculated that the $\beta$-factor increased substantially  (from $\sim$20\% to $\sim$60\%) by increasing $n_s$ from 1.0 to 3.5. In this study, the gap mode is determined by the gap between the metal nanoparticle and the dielectric substrate, rather than the metallic dimer gap in our study. To the best of our knowledge, the effect of the substrate on the $\beta$-factor (also referred to as quantum efficiency), and more importantly, the directionality of the $\beta$-factor has not been demonstrated theoretically or experimentally for metallic dimer structures; however, a recent paper by Zhou \textit{et. al.}\cite{zhang_dipolar_2019} has demonstrated for a single nanorod (dielectric or metallic), the $\beta$-factor is directed predominantly into the higher-index medium if placed on a substrate, supporting our results.  

%\red{no other related works?}

In an attempt to increase the upward directionality of the $\beta$-factor, a 5~nm spacer layer with $n=1.5$ was included between the GaAs substrate and gold dimer. We found that the total $\beta$-factor is increased over the whole broadband spectrum, most notably increasing $\beta$ from 20\% for modes II-IV to over 50\%. The directionality is also slightly increased in the upward direction, but only to less than 10\% of the total. Also, the QNMs are substantially blue-shifted due to the effect of the effective-index of the glass and GaAs heterostructure.

Another potential way to  enhance the upward $\beta$-factor is to include a gold reflector below the substrate (e.g., 100~nm below). We also carried out such calculations and the upper (lower) $\beta$-factor for the first mode is approximately 16\% (40\%), whereas the original structure produced 3\% (71\%), resulting in a change in the ratio of upward-to-downward $\beta$ from ${3}/{71}\approx 0.04$ to ${16}/{40}=\approx 0.40$. However, that the total $\beta$-factor decreases over the entire broadband spectrum (from 74\% to 56\% at mode I and from 6-20\% to less than 3\% over modes II-IV), and the PF remains approximately the same -- note that the modal properties of the resonances were mainly unchanged from those in Fig.~\ref{fig:ns_3.5}. 

One could also attempt to increase the beta factor
by using $\beta$-enhancing hetero-structures and designs \cite{lee_planar_2011, zhang_dipolar_2019, wiecha_finite_2019} %\red{add some related refs}
or microlenses \cite{ee_enhancement_2007}%\red{add some related refs}
, but such designs
are outside the scope of this paper, and they would
be quite challenging to exploit over such a wide
spectral range.

\section{Conclusions}
\label{sec:conclusions}

We have employed a  QNM technique  to study the main resonant modes and complex
frequencies of a gold bowtie dimer with (without) a substrate. The effects of dimer gap size, dipole position, as well as substrate index of refraction were studied for the PF and $\beta$-factor with a dipole oriented along the axis of the dimer. As expected, larger PFs were obtained for smaller gaps with losses to the total $\beta$-factor; however, the total $\beta$-factor, even for small gaps of 0.5-2~nm, remained above 50-60\% which is quite remarkable for an un-optimized design. The $\beta$-factor was further examined by splitting it into its upper and lower hemisphere contributions. Two unexpected observations were made for the inclusion of the substrate of $n_s$=1.2: first, the second mode in the system `split' into two distinctly different modes with different electric field profiles, and the poles of these two modes moved further from each other (in complex frequency space) as the index of the substrate increased. Second, the $\beta$-factor un-intuitively became increasingly directed in the downward (into the substrate) direction as the index increased. This effect can be explained by examining the electric field profiles at the pole, noting how the electric field is `pulled' into the substrate compared to the no-substrate case.  These results are not currently discussed in the literature, and could have possibly important consequences for how the photons in such systems are collected for optimal $\beta$-factor output. Further work includes modelling the structure for optimal properties in PF and $\beta$-factor, including spacer layers, distributed Bragg mirror substrates, and geometrical factors in the bowtie design.
%\blue{update depending upon final graphs}.  

\acknowledgements

This work was supported by the Natural Sciences and Engineering Research Council of Canada, the Canadian Foundation for Innovation and Queen's University.

\bibliography{refs}
\end{document}